# Influence of chemical vapor deposition conditions on N incorporation ratio on vicinal 4H-SiC(000−1) surface: Ab Initio-based approach

Wataru Ota[1], Akira Kusaba[1, 2], Yoshihiro Kangawa[1, 2, *]
Keisuke Kurashima[3], Ichiro Mizushima[3], and Kenji Shiraishi[4]

[1] Interdisciplinary Graduate School of Engineering Sciences, Kyushu University, Fukuoka 816-8580, Japan
[2] Research Institute for Applied Mechanics, Kyushu University, Fukuoka 816-8580, Japan
[3] NuFlare Technology Inc., Yokohama 235-8522, Japan
[4] Center for Innovative Integrated Electronic Systems, Tohoku University, Sendai 980-8572, Japan
*Corresponding author: kangawa@riam.kyushu-u.ac.jp

Abstract:
In experimental studies, increasing the ratio of source gases ($C_3H_8$/$SiH_4$; C/Si ratio) for chemical vapor deposition (CVD) causes the growth rate to increase monotonically, and nitrogen incorporation tends to decrease accordingly; however, a unique phenomenon has been observed in which nitrogen incorporation increases discontinuously in the range of C/Si ratio between 1.0 and 1.5. This phenomenon is specific to the vicinal 4H-SiC(000−1) C-face and is not observed on the vicinal 4H-SiC(0001) Si-face. In this study, N incorporation behavior on a vicinal C-face during CVD is investigated using an ab initio-based approach. The calculation results suggest that when C/Si ratio is less than 1.0, the Si-terminated step edge is stable, whereas when C/Si ratio exceeds 1.0, the existence probability of the C-H-terminated step edge, where the N substitution energy is lower than in the former case, increases sharply. This change in the step-edge state across C/Si = 1.0 is thought to be the cause of the discontinuous increase in N incorporation.

## 1. Introduction

Silicon carbide (SiC) is a wide-bandgap semiconductor with excellent properties, and 4H-SiC is widely used in power devices. 4H-SiC for power device applications is primarily grown by chemical vapor deposition (CVD), which involves multiple elementary processes, including gas-phase reactions [1-3], precursor adsorption on reconstructed surfaces [4-7], surface migration, partial desorption, and incorporation at step edges [8]. $N_2$ is commonly used for n-type doping, and N incorporation on the vicinal 4H-SiC(0001) Si-face has also been investigated by first-principles calculations [9].

In contrast, the mechanism of N incorporation on the C-face and its atomistic understanding remain unclear. Furthermore, experimental studies have observed unique nitrogen-incorporation behavior on the 4H-SiC(000-1) C-face. **Figures 1(a)** and **1(b)** show the growth rate and N doping efficiency as a function of the C/Si ratio in the supply gas, respectively. The CVD growth was performed on a 6” wafer using EPIREVO$^{TM}$ S8, which is a trademark of NuFlare Technology, Inc., under the conditions of $SiH_4$ flow rate of 360 sccm, growth temperature of 1625 $^{o}$C and wafer rotation speed of 600 rpm. The growth rate shows no significant difference between the Si-face and the C-face; both increase monotonically as the C/Si ratio increases (**Fig. 1(a)**). On the other hand, while N-doping efficiency tends to decrease as the C/Si ratio increases, a unique phenomenon occurs on the C-face: it increases discontinuously in the range C/Si = 1.0–1.5 (**Fig. 1(b)**). Although the experimental conditions differ, a similar phenomenon shown in **Fig. 1(b)**—in which N-doping efficiency changes discontinuously at a C/Si ratio of approximately 1.1—has also been observed [10-13]. While one report [14] discusses the relationship between N-doping efficiency and SiC growth kinetics, it found no discontinuous change in the growth rate (**Fig. 1(a)**); therefore, this is considered a separate phenomenon. Generally, it is believed that N, which substitutes for a C site, is more easily incorporated into the Si-face—where the topmost C has a single bond with a lower-layer Si atom—than into the C-face, where the topmost C has triple bonds with lower-layer Si atoms [15]. However, the unique phenomenon observed in **Fig. 1(b)** depends on the growth conditions (C/Si ratio) rather than on substrate orientation, i.e., Atomic arrangements of the ideal surfaces. Based on the above discussion, this study investigated, using an ab initio-based approach [16-20], the correlation between the bonding state of the 4H-SiC(000-1) C-face—specifically, the dependence of the growth condition (C/Si ratio) on surface reconstruction—and N-doping efficiency. Here, we also include step-edge structures in the analyses to discuss realistic growth mechanisms.

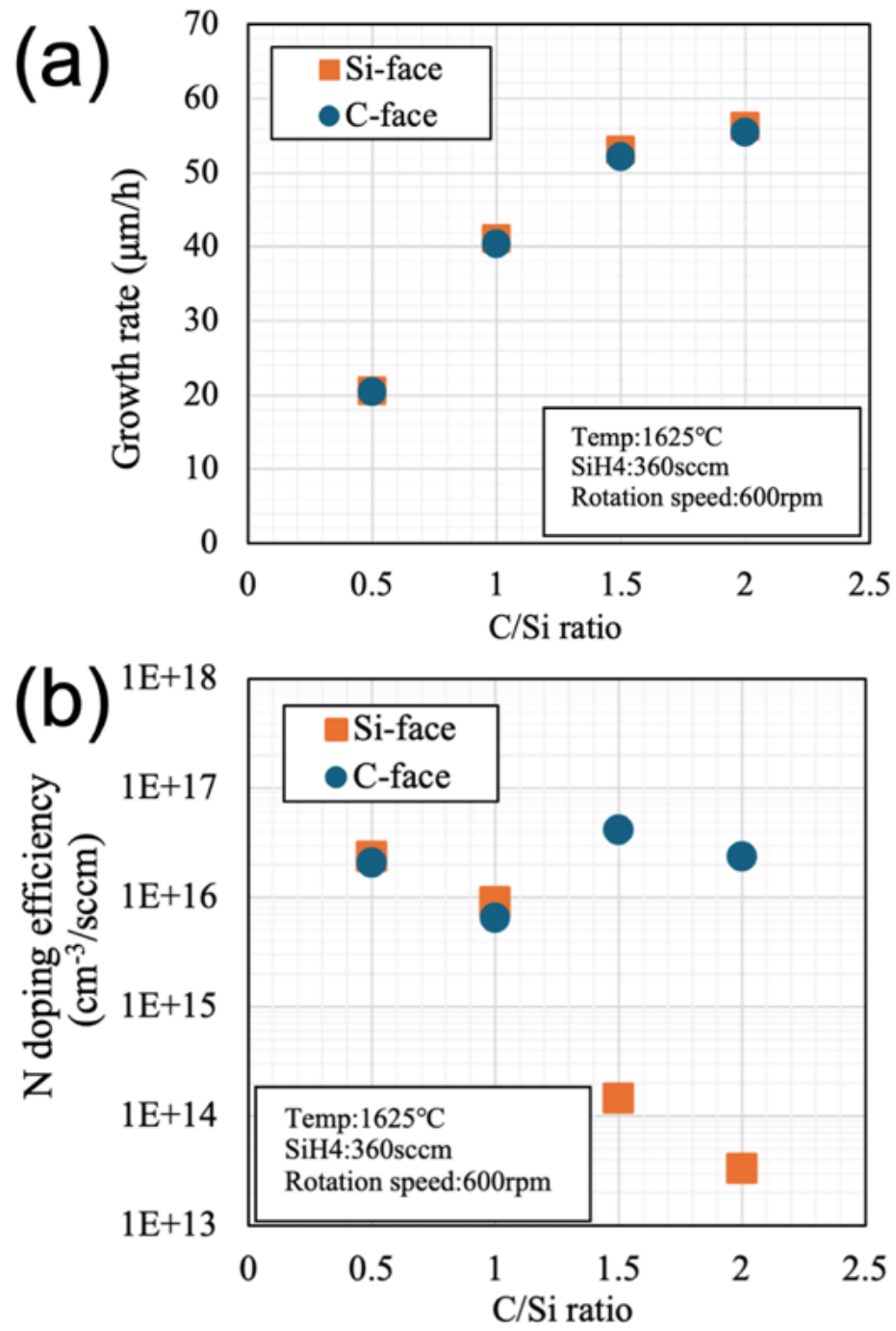

**Fig. 1** Dependence on C/Si ratio in supply gas of (a) growth rate and (b) N doping efficiency, obtained by averaging within a 6” wafer, grown using EPIREVO™ S8. EPIREVO is a trademark of NuFlare Technology, Inc.

## 2. Computational methods

**Figures 2(a)** and **2(b)** show the 4H-SiC(11−20) side view and (000−1) top view of a 2 × 2 slab model drawn by the visualization software VESTA [21]. In **Fig. 2(a)**, the 20 Å-thick vacuum layer above the surface has been omitted. Blue and brown atoms represent silicon and carbon, respectively. Four adsorption sites (i-iv) exist on the topmost carbon atoms. The slab is 8 monolayers (MLs) thick along [000−1]. The atomic positions of one ML of SiC at the bottom and the terminated hydrogen (pink atoms) were fixed to mimic the bulk state, and the remaining atoms were relaxed. We used the Vienna Ab-initio Simulation Package (VASP) [22] within the density functional theory (DFT) framework with the PBE generalized gradient approximation [23] for the exchange-correlation functional. All calculations were performed using projector-augmented wave (PAW) [24, 25] pseudopotentials with a plane-wave cutoff energy of 500 eV, as implemented in VASP. The number of Monhorst-Pack k-point sampling [26] used in this model was 3 × 3 × 1. This level of accuracy is sufficient to discuss relative stability in geometry optimization; we set the energy convergence threshold to $1.0 \times 10^{-6}$ eV and the maximum force component to $1.0 \times 10^{-3}$ eV/Å.

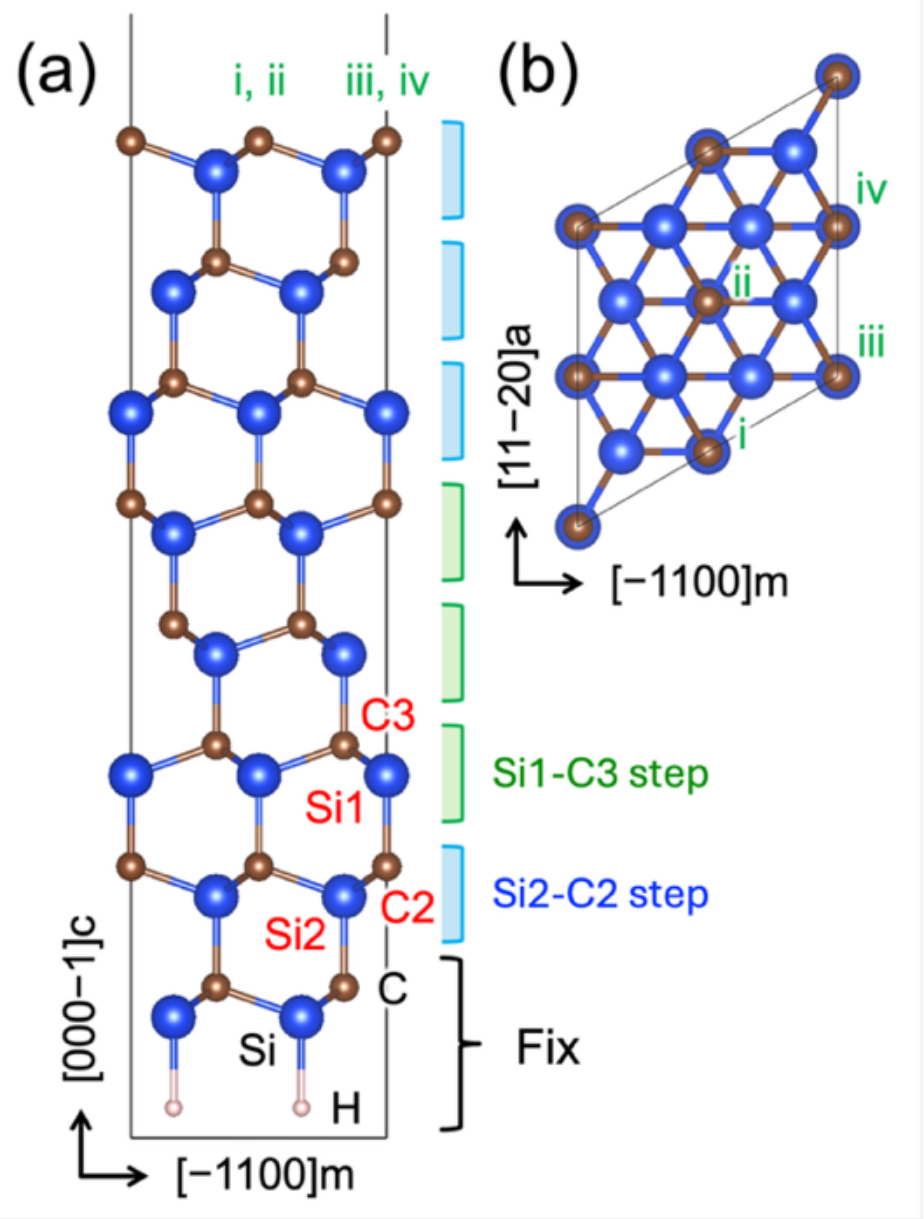

**Fig. 2** Schematic of slab model; (a) (11−20) side view, and (b) (000−1) top view. The 20Å-thick vacuum layer has been omitted. Blue and brown atoms show silicon and carbon atoms, respectively. The bottom surface was terminated by hydrogen (colored pink) to mimic the bulk state. There are four adsorption sites (i-iv) on the topmost carbon atoms. When the surface is tilted toward [1−100], there is a possibility of two sets of step edges: (1) Si2 or C2 step edges, or (2) Si1 or C3 step edges. Here, Si2 means a Si atom with two dangling bonds. The same applies to other notations.

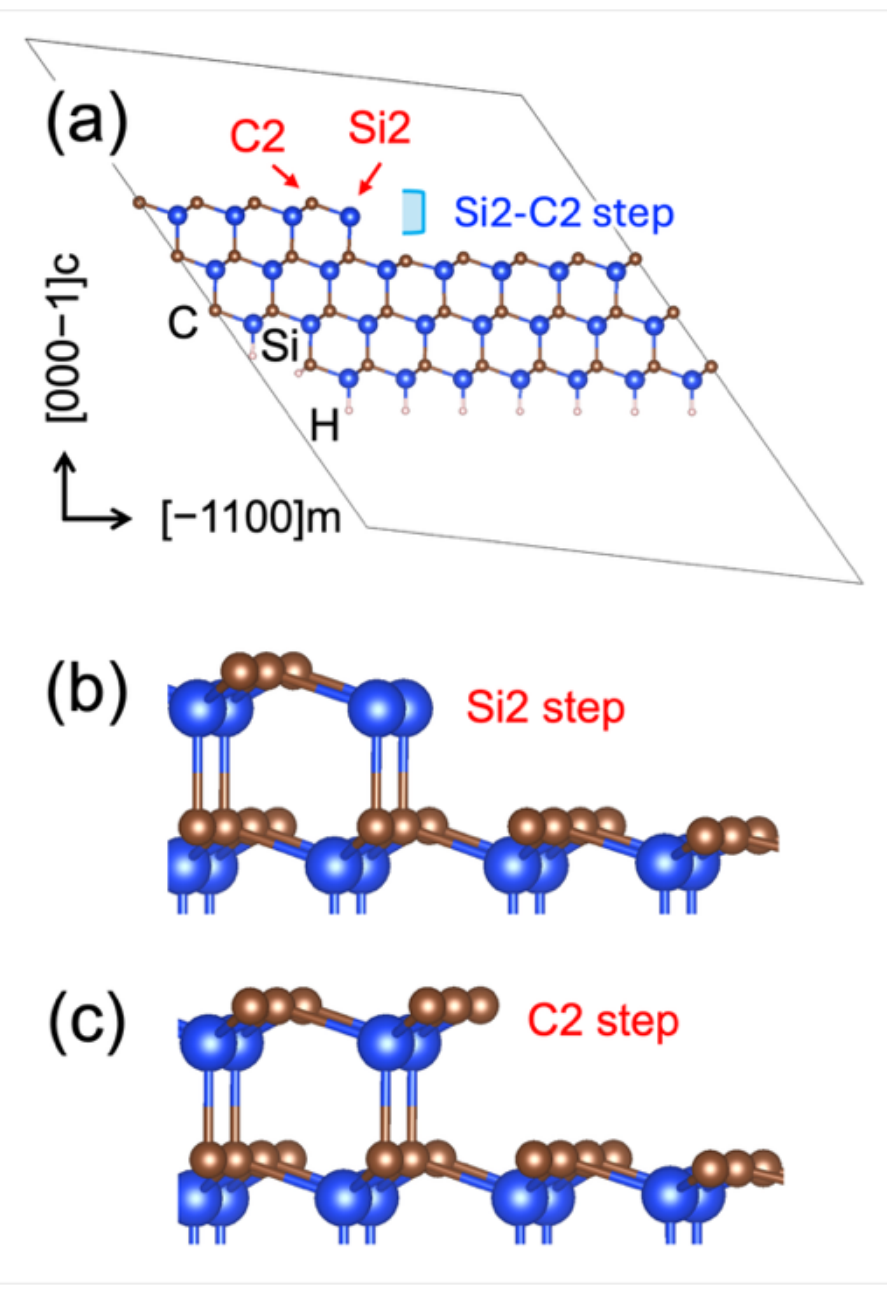

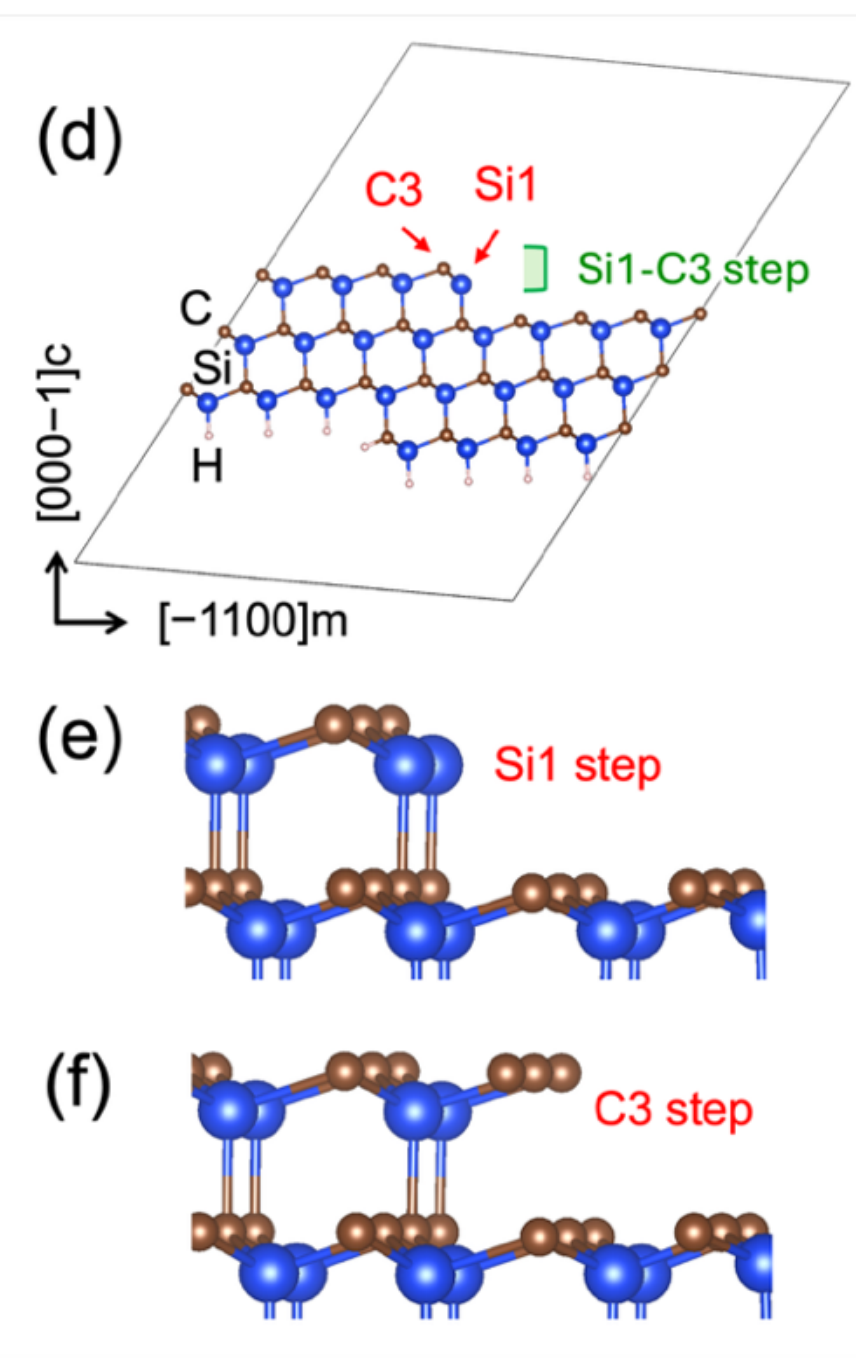

**Fig. 3** (a) and (d) are schematics of the vicinal slab model. Blue and brown atoms represent silicon and carbon, respectively. The bottom surface was terminated with hydrogen (colored pink) to mimic the bulk state. On the [−1100]-off 4H-SiC(000−1) substrate, two types of step edge sets appear: (b) Si2 or (c) C2 step edge and (e) Si1 or (f) C3 step edge. Here, Si2 means a Si atom with two dangling bonds. The same applies to other notations. In (c) and (f), three carbon atoms are seen along the step edge. One of them is duplicated due to the periodic boundary conditions.

When the surface is tilted toward [−1100], there is a possibility of two sets of step edges: (1) Si2 or C2 step edges, or (2) Si1 or C3 step edges (See **Fig. 2** and **3**). Here, Si2 means a Si atom with two dangling bonds at the step edge. The same applies to other notations. **Figures 3(a)** and **3(d)** show the slab models with Si2-C2 and Si1-C3 step edges, respectively. The slab models are 2 ML thick along [11−20], with two Si or C atoms at the step edge. At the C2 step edge (**Fig. 3(c)**) and the C3 step edge (**Fig. 3(f)**), three C atoms are seen at the step edge, but one of them is a duplicate caused by the periodic boundary conditions. In the ab initio calculations for these vicinal slab models with a surface step, we used 1 × 3 × 1 k-point sampling; all other conditions were the same as described above.

The adsorption and desorption behavior of precursors on a CVD growth surface can be analyzed based on the sign of the adsorption energy $E_{\mathrm{ad}}$ in the following equation:

$$E_{\mathrm{ad}} = E_{\mathrm{adsorption}} - \{E_{\mathrm{desorption}} + n_i\mu_i^{gas}\}, \tag{1}$$

$$\mu_i^{gas} = E_i^{gas} + \mu_i, \tag{2}$$

where $E_{\mathrm{adsorption}}$ and $E_{\mathrm{desorption}}$ are the total energies of the surface slab model with and without adsorbed atom/molecules $i$; $E_i^{gas}$ is the total energy of an atom/molecule $i$ in the gas phase; $n_i$ is the number of adsorbed atom/molecules $i$ on the surface. The chemical potential $\mu_i$ for an atom/molecule $i$ is expressed as a function of partial pressure $p_i$ and growth temperature $T$ in statistical thermodynamics. For details, see the literature [16, 19, 20]. In this study, we used the Thermochemistry Module of the Atomic Simulation Environment (ASE) [27] to calculate the chemical potentials of precursors. However, to ensure an appropriate comparison, we excluded zero-point vibrations for both the precursors and the crystals. As shown in **Fig. 4**, when $E_{\mathrm{ad}}$ is negative, adsorption predominates, and when it is positive, desorption predominates. Using this adsorption-desorption concept, the Gibbs free energy $G$ of the surface can be calculated as follows [28].

$$G = E_{\mathrm{reconstruction}} - \left(E_{\mathrm{ideal}} + \textstyle\sum_i n_i\mu_i^{gas}\right), \tag{3}$$

where $E_{\mathrm{reconstruction}}$ and $E_{\mathrm{ideal}}$ are the total energies of reconstructed and ideal surfaces, respectively. Since $\mu_i^{gas}$ is a function of $p_i$ and $T$, we can plot the $p$−$T$ phase diagram using equation (3).

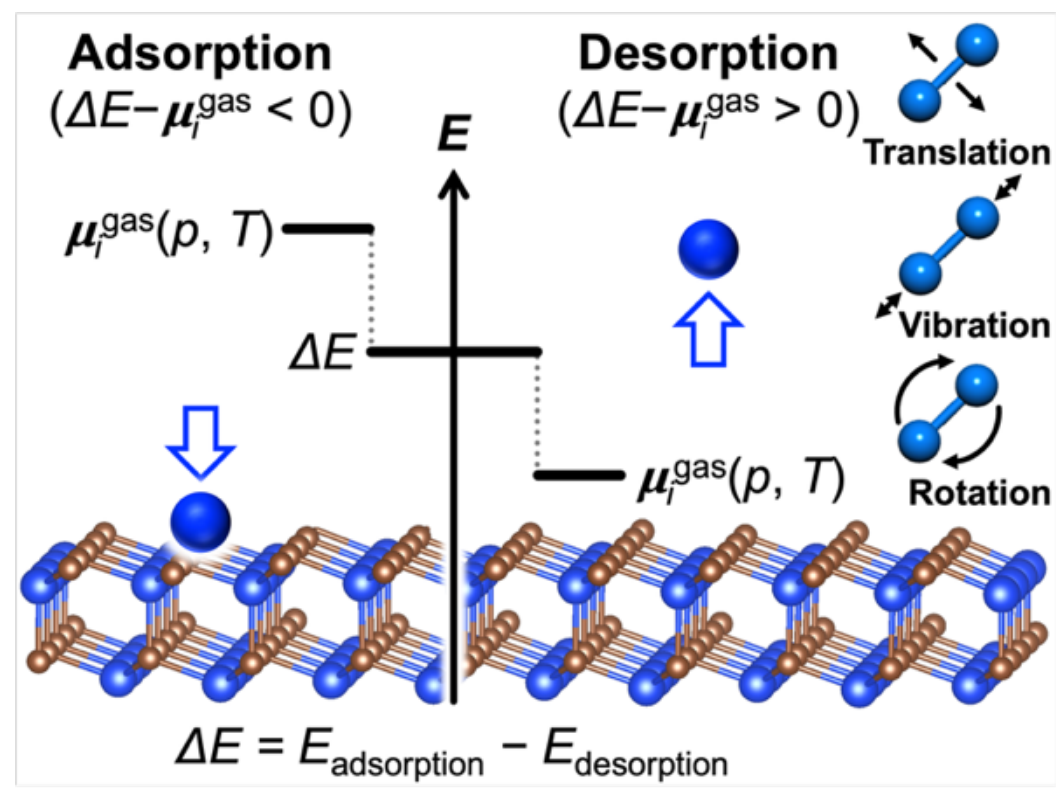

**Fig. 4** Conceptual figure of an ab initio-based approach to analyze adsorption-desorption behavior of precursors in chemical vapor deposition. If the chemical potential of the precursor in the gas phase, $\mu^{gas}$, is higher than the energy difference, $\Delta E$ (= $E_{adsorption} - E_{desorption}$), the precursor is stable on the surface. In the opposite relationship, the precursor is stable in the gas phase. The chemical potential $\mu^{gas}$ is expressed as a function of partial pressure, $p$, and temperature, $T$, using the partition functions for translational, vibrational, and rotational motion.

## 3. Results and discussion

To compare the experimental results above, we performed analyses based on the following CVD conditions using $SiH_4$ and $C_3H_8$ as source gases: C/Si = $1.0\times10^{-2}$–$1.0\times10^{+2}$, Cl/Si = 10, Si/H = $2.36\times10^{-3}$, partial pressure of $H_2$ dilute gas $p_{H2}$ is 0.25 atm, and $T$ = 1400−1800 °C. Chokawa et al. performed thermodynamic analyses and reported that, under the above CVD conditions with halides, the major precursors are $SiCl_2$ and $C_2H_2$ [7]. In the present study, we analyzed the surface phase diagram assuming that all input source gases convert into $SiCl_2$ and $C_2H_2$ precursors and are supplied to the growth surface.

### 3.1 H or Cl coverage on the terrace

First, we analyzed the coverage of H or Cl—dilute and additional gas elements, respectively—on 4H-SiC(000-1) (C-face) without step structures, i.e., on the terrace. On the terrace, within a 2 × 2 unit cell, four H or Cl adsorption sites—labeled i–iv—exist, as shown in **Fig. 2**. Here, a model in which an H atom is adsorbed at one of the four adsorption sites is denoted [H1]; a model in which four Cl atoms are adsorbed at the four adsorption sites is denoted [Cl4]; and the ideal surface without adatoms is denoted [0]. For adsorbed $H_2$ or HCl gases, the surface energy is given by the following equation, derived from Equation (3).

$$G = E_{\text{reconstruction}} - \left\{E_{\text{ideal}} + n_{\text{H}}\frac{1}{2}\mu_{\text{H2}}^{gas} + n_{\text{Cl}}\left(\mu_{\text{HCl}}^{gas} - \frac{1}{2}\mu_{\text{H2}}^{gas}\right)\right\}. \quad (4)$$

Under experimental conditions—specifically, a partial pressure of $H_2$ of 0.25 and a partial pressure of HCl of $6.0 \times 10^{-3}$ atm—the temperature dependence of the surface energy is as shown in **Fig. 5(a)**. In Equation (4), the surface energy of the ideal surface—that is, the surface energy of [0]—is 0.0 eV. At temperatures between 1400 and 1800 °C under the above gaseous conditions, [H4] is the most stable. Assuming a Boltzmann distribution, the existence probability $P_i$ of surface reconstruction $i$ can be calculated using the

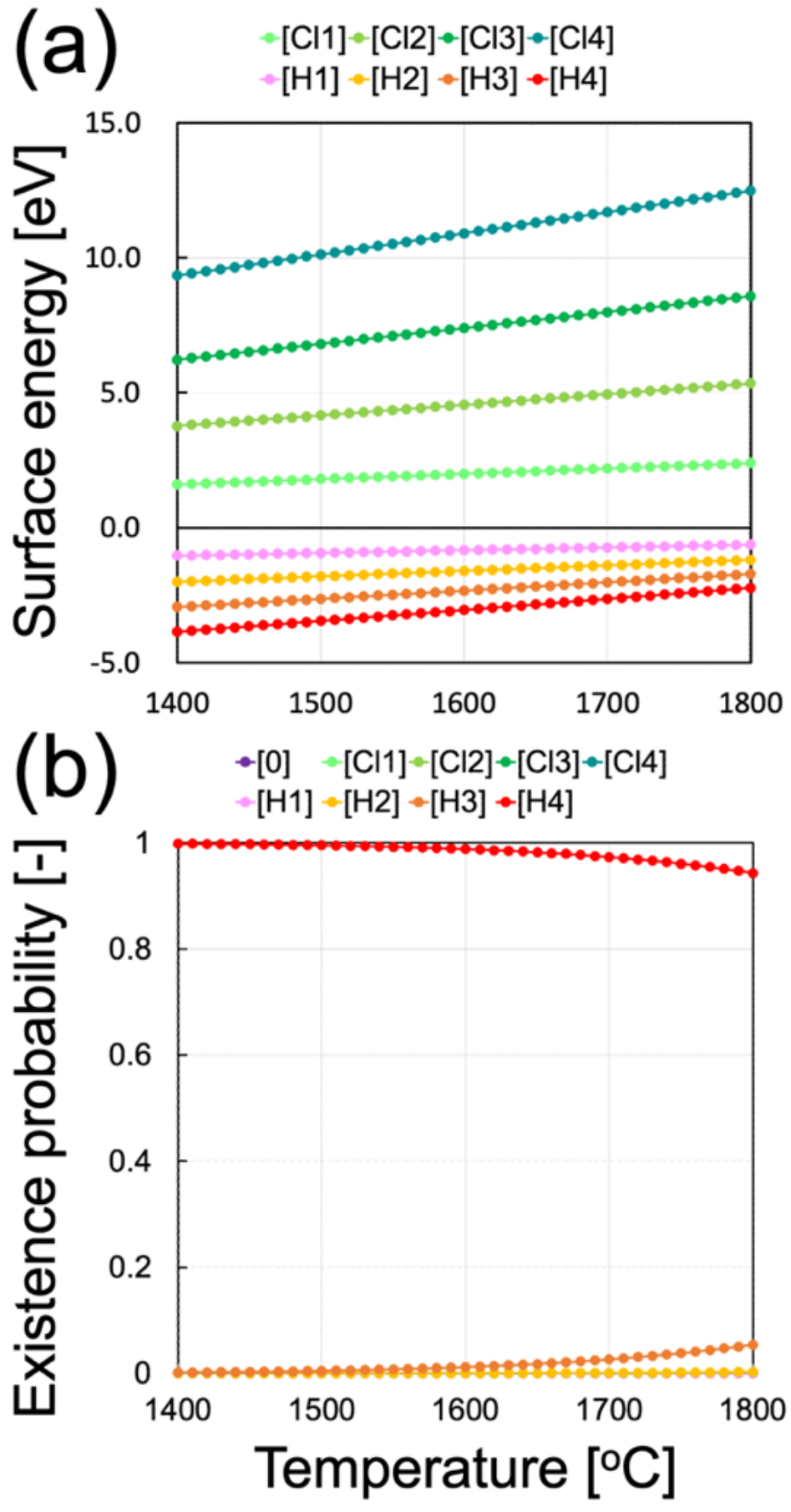

**Fig. 5** (a) Surface energy of the 2x2 slab model (see Fig. 1) as a function of temperature. [Cl1] indicates that one of the four surface adsorption sites (i-iv) is terminated with a Cl atom. The same applies to other notations. (b) The existence probability of each terminated surface as a function of temperature. The existence probability of the [H4] surface exceeds 0.94 in the temperature range between 1400 and 1800 °C.

following equation.

$$P_i = \frac{exp\left(\frac{-G_i}{k_B T}\right)}{\sum_j exp\left(\frac{-G_j}{k_B T}\right)}, \quad (5)$$

where $k_B$ is Boltzmann's constant. The summation over $j$ is performed for all surface reconstructions: [0], [Cl1]–[Cl4], and [H1]–[H4]. **Fig. 5(b)** shows the calculated existence probability $P_i$ as a function of temperature. Within this temperature range (1400–1800 °C), the existence probability of [H4] being present exceeds 94%, suggesting that the adsorption sites on the terrace are almost entirely covered/terminated by H. The H coverage of ~100% on the C-face terrace is higher than that reported by Kimura et al. [8] for the Si-face terrace (50–60% at $p_{H2}$ = 0.25 atm, $T$ = 1625 °C). This result is reasonable considering that a relatively strong bond forms between the topmost C and H on the C-face, whereas a relatively weak bond forms between the topmost Si and H on the Si-face.

## 3.2 Step edge termination

Next, we calculated H- and Cl-termination at the step edge on the [−1100]-off 4H-SiC(000−1), where the terrace was entirely covered by H. Specifically, we analyzed the H or Cl termination of the step edge using a vicinal model in which H was adsorbed onto the dangling bond of the topmost C atom shown in **Figs. 3(a)** and **3(d)**. However, we assumed that the C atoms at the bottom of the step were H-free, as it is easy to infer that they would make bonds with C2 or C3 atoms at the step edge (See **Figs. 3(c)** and **3(f)**). Please refer to the section below for details on the relaxed structure. As mentioned above, because there are two Si or C atoms along the step edge, the following H- or Cl-termination models are possible: [0], [H1], [H2], [Cl1], and [Cl2]. Here, we analyzed the stability of the step-edge structure using the following equation derived from Eq. (3), using $SiCl_2$ and $C_2H_2$ as precursors.

$$G = E_{\mathrm{reconstruction}} - \Big\{E_{\mathrm{ideal}} + n_{\mathrm{H}}\frac{1}{2}\mu_{\mathrm{H2}}^{gas} + n_{\mathrm{Cl}}\left(\mu_{\mathrm{HCl}}^{gas} - \frac{1}{2}\mu_{\mathrm{H2}}^{gas}\right) + n_{\mathrm{Si}}\left(\mu_{\mathrm{SiCl2}}^{gas} - 2\mu_{\mathrm{HCl}}^{gas} - \mu_{\mathrm{H2}}^{gas}\right) + n_{\mathrm{C}}\frac{1}{2}\left(\mu_{\mathrm{C2H2}}^{gas} - \mu_{\mathrm{H2}}^{gas}\right)\Big\}. \quad (6)$$

**Figures 6(a)** and **6(b)** show the stability of step structures for the Si2-C2 step edge and for the Si1-C3 step edge, respectively, as a function of C/Si ratio and temperature. The gaseous pressure conditions for the calculation are described in the figure caption. In the case of the Si2-C2 step edge (**Fig. 6(a)**), Si2[0] is stable under low C/Si ratio and high-temperature conditions, while C2[H2] is stable under high C/Si ratio and low-temperature conditions. Here, Si2[0] implies that two Si2 atoms at the step edge are not terminated with H and/or Cl; C2[H2] implies that two C2 atoms at the step edge are terminated with an H adatom, respectively. In the case of the Si1-C3 step edge (**Fig. 6(b)**), a Si1[0] step edge would appear.

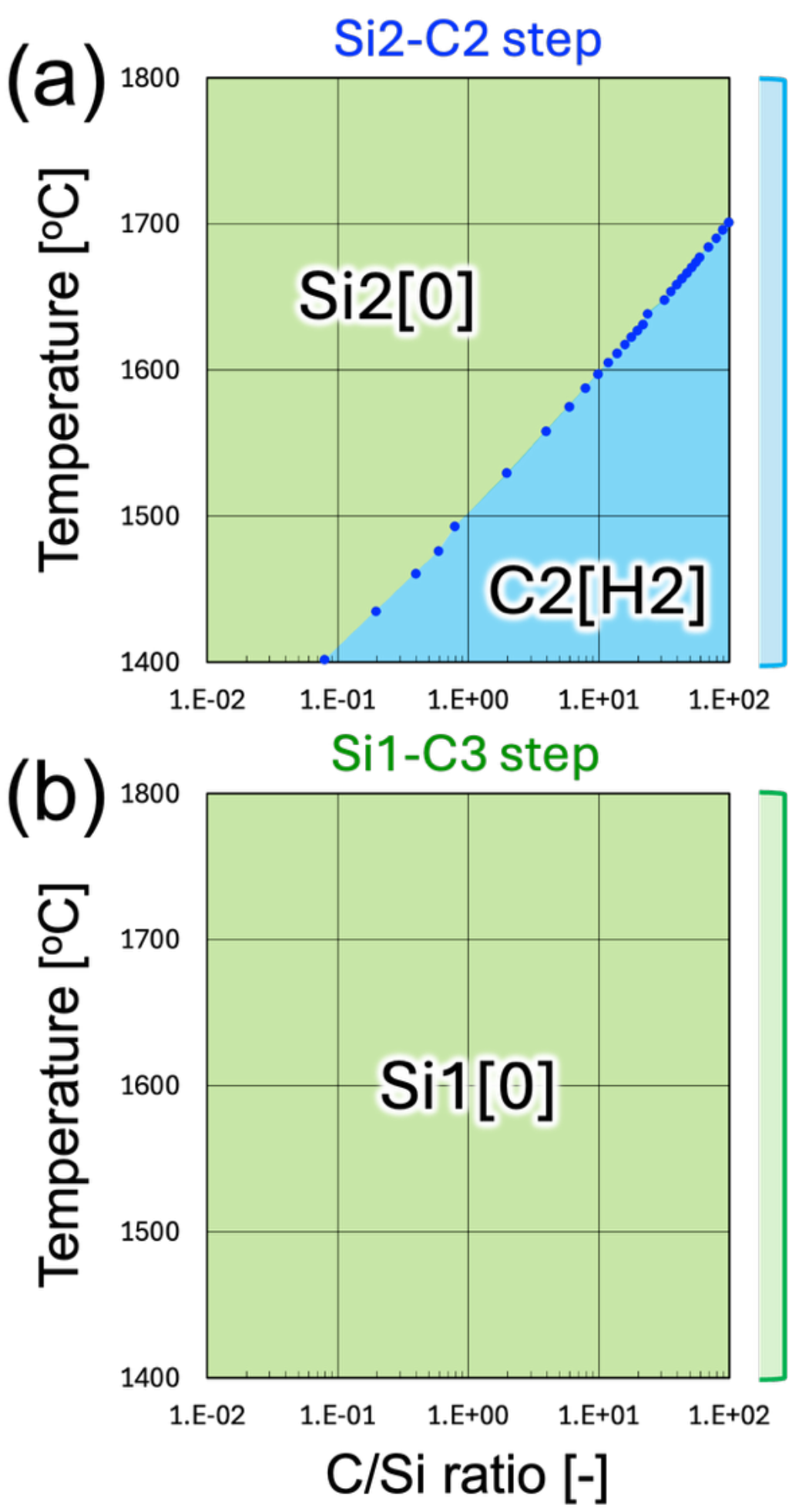

**Fig. 6** Stability of step edge structure. For the Si2-C2 step edge set, the Si2[0] step edge is stable under high-temperature, low C/Si ratio conditions (a). Here, Si2[0] indicates that the two Si2 atoms at the step edge are not terminated with H or Cl. On the other hand, the C2[H2] step edge becomes stable under low-temperature, high C/Si ratio conditions. Here, C2[H2] indicates that both of the two C2 atoms at the step edge are terminated with H. For the Si1-C3 step edge set, the Si1[0] step edge is stable across the entire condition range. The calculation conditions are Cl/Si = 10, Si/H = $2.36\times10^{-3}$, and $p_{H2}$ = 0.25 atm.

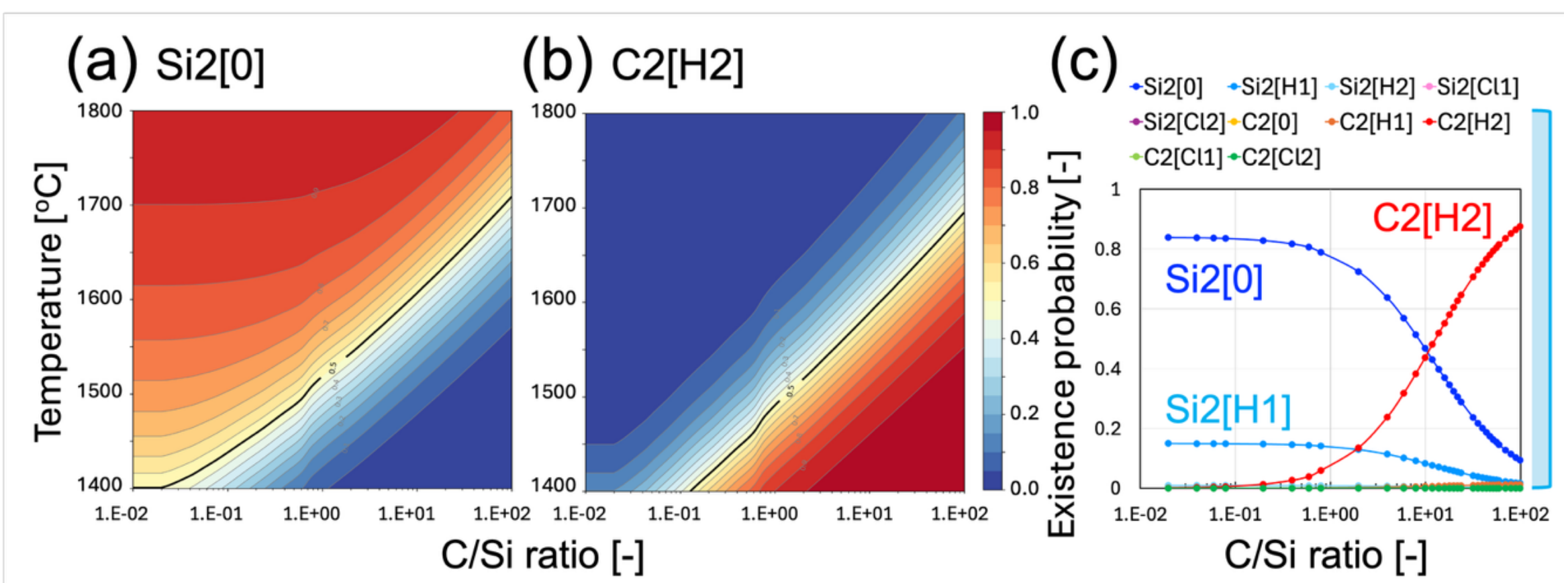

**Fig. 7** Contour map of the existence probability of (a) Si2[0] and (b) C2[H2] for the Si2-C2 step edge set. For information on the notation for C2[H2] and Si2[0], please refer to the caption for Fig. 6. (c) The existence probability of each terminated step edge as a function of C/Si ratio at 1600 ºC. Si2[0] is the most stable under low C/Si ratio, and C2[H2] under high C/Si ratio. When C/Si ratio exceeds 1.0, the existence probability of C2[H2] occurring increases sharply. The calculation conditions are Cl/Si = 10, Si/H = $2.36\times10^{-3}$, and $p_{H2}$ = 0.25 atm.

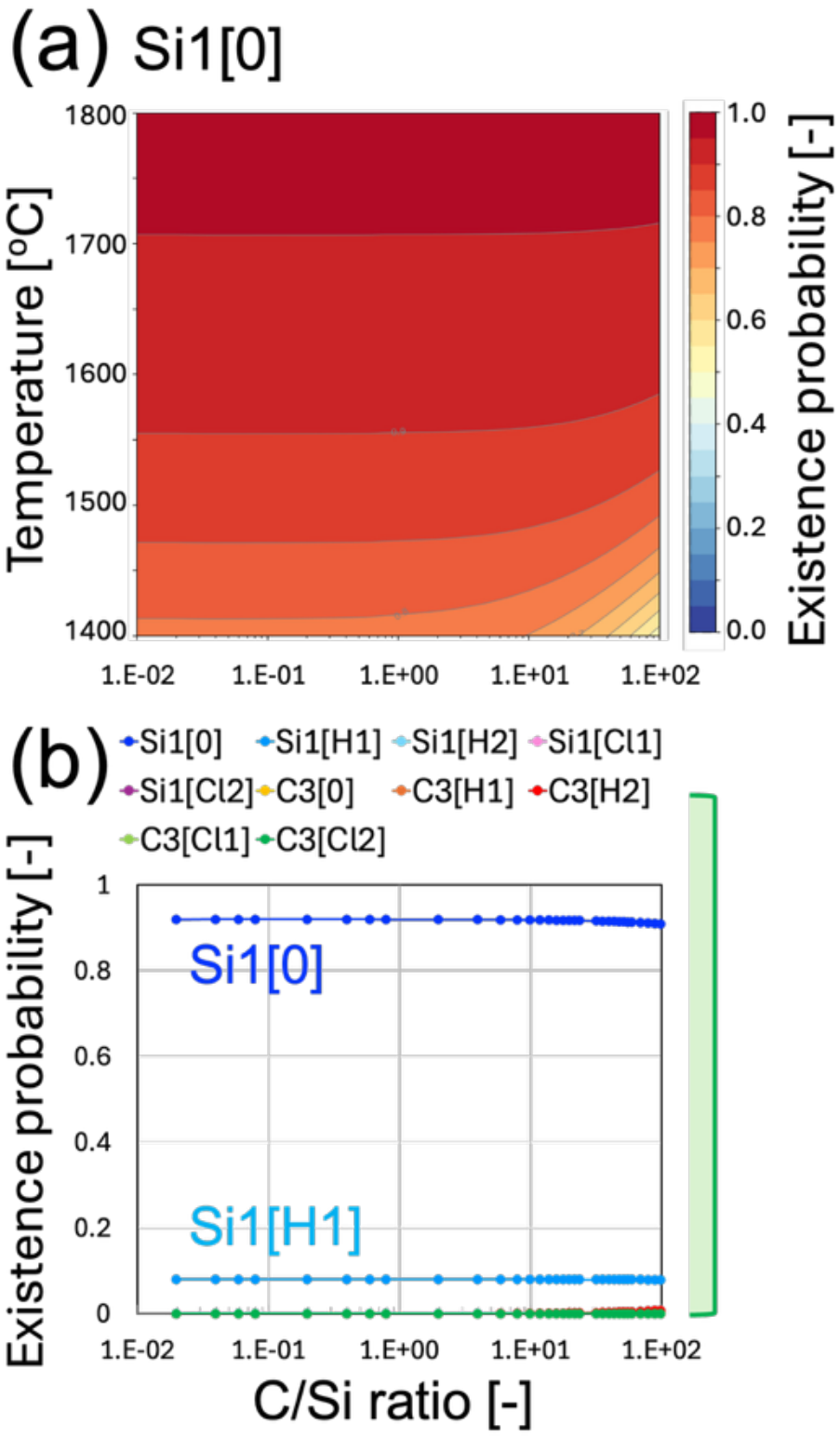

**Fig. 8 (a)** Contour map of the existence probability of Si1[0] for the Si1-C3 step edge set. Here, Si1[0] indicates that the two Si1 atoms at the step edge are not terminated with H or Cl. (b) The existence probability of each terminated step edge as a function of C/Si ratio at 1600 ºC. Si1[0] is the most stable across the entire condition range. The calculation conditions are Cl/Si = 10, Si/H = $2.36\times10^{-3}$, and $p_{H2}$ = 0.25 atm.

**Figures 7(a)** and **7(b)** show the contour map of the existence probability of Si2[0] and C2[H2] calculated by equation (5), respectively. The summation over $j$ is performed for all step-edge structures: [0], [Cl1], [Cl2], [H1], and [H2] for Si2 and C2 step edges, respectively. Corresponding to **Fig. 6(a)**, the existence probability of Si2[0] is large under low C/Si ratio and high-temperature conditions, while that of C2[H2] is larger under high C/Si ratio and low-temperature conditions. **Figure 7(c)** shows the existence probability of each Si2-C2 step-edge structure at 1600 °C. It can be seen that when the C/Si ratio exceeds 1.0, the existence probability of the stable Si2[0] and the metastable Si2[H1] decreases, while that of the C2[H2] increases sharply.

**Figure 8(a)** shows the contour map of the existence probability of Si1[0] calculated by equation (5). The summation over $j$ is performed for all step-edge structures: [0], [Cl1], [Cl2], [H1], and [H2] for Si1 and C3 step edges, respectively. Corresponding to **Fig. 6(b)**, within the range of the analysis conditions, the existence probability of Si1[0] is the highest among the calculation models. **Figure 8(b)** shows the existence probability of each Si1-C3 step-edge structure at 1600 °C. In the range of C/Si ratio from $1.0\times10^{-2}$ to $1.0\times10^{+2}$, the existence probability of the most stable Si1[0] exceeds 90%, and that of the metastable Si1[H1] is approximately 8%.

### 3.3 N incorporation at the step edge

Based on the above discussion, for both the Si2-C2 and Si1-C3 step edges, the most stable and metastable structures were [0], [H1], or [H2], and the existence probability of a Cl-terminated step edge series was negligible. Therefore, we analyzed the N substitution energy at the step edge for the H-termination series. The N substitution energy when using $N_2$ doping gas is given by the following equation.

*for N substituting Si*:

$$E_{\text{N substitute Si}} = \{E_{\text{N}} + (E_{\text{SiCl2}}^{gas} - 2E_{\text{HCl}}^{gas} + E_{\text{H2}}^{gas})\} - \{E_{\text{Si}} + \tfrac{1}{2}E_{\text{N2}}^{gas}\}, \quad (7)$$

*for N substituting C*:

$$E_{\text{N substitute C}} = \{E_{\text{N}} + \tfrac{1}{2}(E_{\text{C2H2}}^{gas} - E_{\text{H2}}^{gas})\} - \{E_{\text{C}} + \tfrac{1}{2}E_{\text{N2}}^{gas}\}, \quad (8)$$

where $E_{\text{N}}$ and $E_{\text{Si}}$ ($E_{\text{C}}$) are the total energy after and before N substituting Si (C), respectively. We used Equation (7) for Si2 and Si1 step edges, and Equation (8) for C2 and C3 step edges. It is noted that general equations (7) and (8) represent the N-substitution energy at absolute zero (0 K). **Table 1** shows the computed N substitution energy at each step edge.

In the case of the Si2-C2 step edge, Si2[0] is the most stable and Si2[H1] is the metastable step edge structure under low C/Si ratio and high-temperature conditions. The N substitution energies at Si2[0] and Si2[H1] are 0.32 and 0.90 eV/atom, respectively. Under high C/Si ratio and low-temperature conditions, C2[H2] is stable, and the N substitution energy at the step edge is 0.15 eV/atom, which is lower than that of Si2[0] and Si2[H1]. **Figures 9(a)** and **9(b)** show the atomic structures before and after N substituting C at the C2[H2] step edge, respectively. Before the N-substitution, the C2 atom at the step edge forms a C-C bond with the C atom at the bottom of the step, and each C atom at the step edge is terminated with H (**Fig. 9(a)**). When an N substitutes one of the C atoms at the step edge, one C-C bond is not replaced by an N-C bond; it would be broken (**Fig. 9(b)**). If an H atom was released from C2[H2], C2[H1] step edge structure appears (**Fig. 9(c)**). An N atom easily substitutes for the step-edge C without H rather than with H (**Fig. 9(d)**). In this case, one C–C bond is replaced by an N–C bond, stabilizing the step-edge structure. Here, the N substitution energy for C at the C2[H1] step edge is –3.01 eV/atom, which is more favorable than that at C2[H2]. In other words, when C/Si ratio exceeds 1.0 and the existence probability of an H-terminated C2 step edge sharply increases, it suggests that the N doping efficiency also increases sharply.

**Table 1** N substitution energy at each step edge. $N_2$ is being considered as a nitrogen dopant.

| Step edge type | N substitution energy |
|---|---|
| C2[0] | −0.79 eV/atom |
| C2[H1] | −3.01 eV/atom |
| C2[H2] | 0.15 eV/atom |
| Si2[0] | 0.32 eV/atom |
| Si2[H1] | 0.90 eV/atom |
| Si2[H2] | 0.91 eV/atom |
| **Step edge type** | **N substitution energy** |
| C3[0] | −1.86 eV/atom |
| C3[H1] | −1.98 eV/atom |
| C3[H2] | 1.89 eV/atom |
| Si1[0] | 2.03 eV/atom |
| Si1[H1] | 2.63 eV/atom |
| Si1[H2] | 2.57 eV/atom |

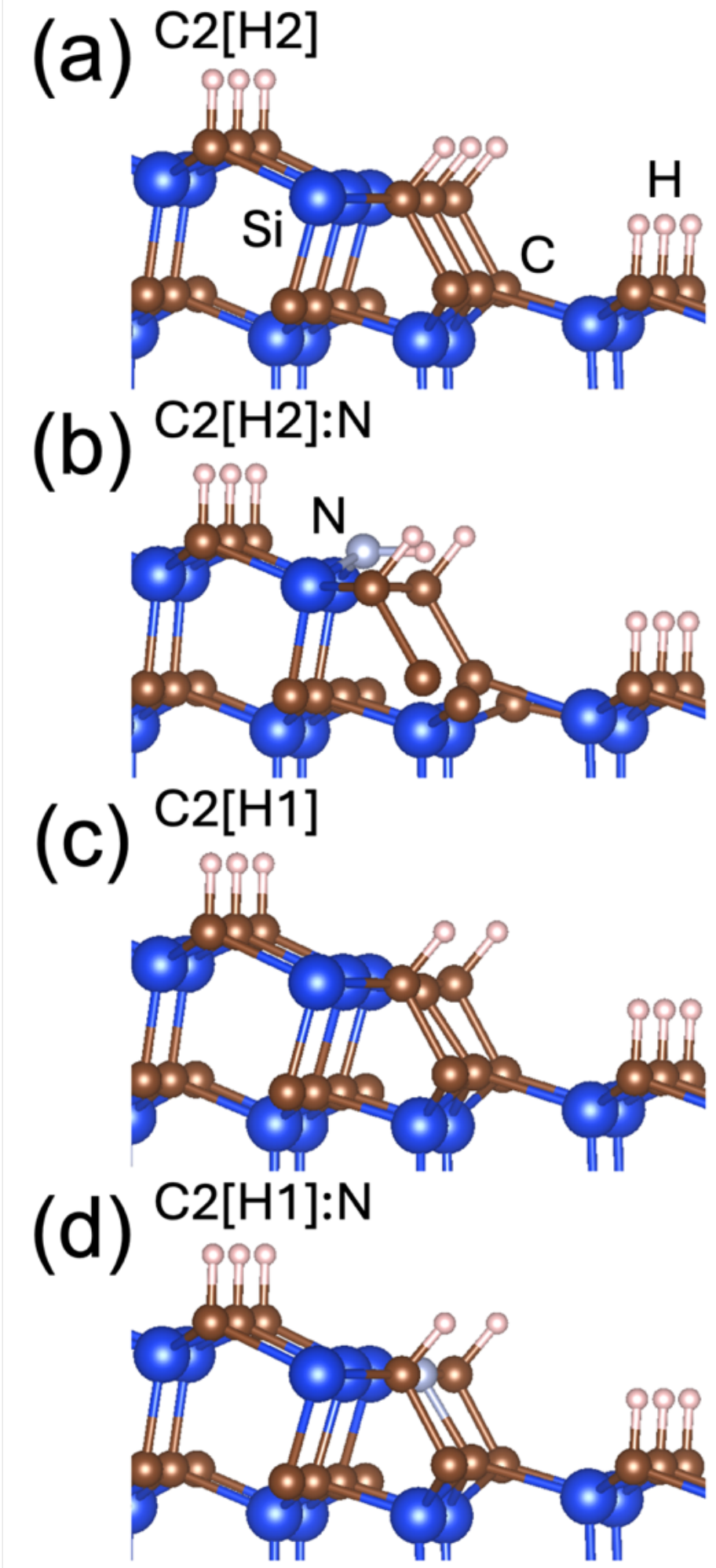

**Fig. 9** Relaxed atomic structures of (a) C2[H2], (b) C2[H2]:N, (c) C2[H1], and (d) C2[H1]:N step edge. In the figures, three C or C-H atoms are seen along the step edge. One of them is duplicated due to the periodic boundary conditions. When a C2 atom at the C2[H2] step edge is substituted by an N atom, one of the two C-N bonds is broken, and the system becomes unstable (b). At the C2[H1] step edge, the C2 atom without H is readily substituted by an N atom without bond breaking (d).

For the Si1-C3 step edge, Si1[0] is the most stable, and Si1[H1] is the metastable step-edge structure under the analyzed conditions. The N substitution energies at Si1[0] and Si1[H1] are 2.03 and 2.63 eV/atom, respectively. They are higher than those for the Si2-C2 step-edge structures. In other words, it is concluded that the N-doping efficiency at the Si1-C3 step edge is lower than that at the Si2-C2 step edge.

Based on the above, it can be concluded that the N doping efficiency in [−1100]-off 4H-SiC is governed by the N incorporation rate at the Si2-C2 step edge, rather than at the Si1-C3 step edge. At the Si2-C2 step edge, a phase transition in the step-edge structure occurs at C/Si ratios of 1.0–1.5, causing a discontinuous change in N-doping efficiency. The calculation results agree well with the experimental results (**Fig. 1(b)**); we show how CVD conditions influence N-doping efficiency in 4H-SiC(000−1) (C-face) with surface steps.

## 4. Conclusions

We experimentally observed a unique phenomenon in which the N-doping efficiency changes discontinuously at a C/Si ratio of 1.0 during CVD growth with additional halides on the 4H-SiC(000−1) (C-face). We found that in 4H-SiC, two types of step edges (Si2-C2 and Si1-C3 step edge sets) appear depending on the stacking sequence. We also found that a structural phase transition occurs near C/Si = 1.0 at one of these edges (the Si2-C2 step edge), and that this transition enhances N substitution. Although conventional first-principles calculations apply only to systems at absolute zero, making comparison with experimental conditions difficult, we obtained physical insights that accurately reproduce the experimental results using a unique ab initio-based approach grounded in quantum mechanics and statistical thermodynamics. This ab initio-based approach can also be applied to analyze CVD processes in other material systems.

Acknowledgments

This work was partially supported by JSPS KAKENHI Grant Number JP24H00432 and Special Projects by the Institute for Molecular Science (IMS program 26-IMS-C120). The computation was carried out using the computer resources offered under the category of General Projects by the Research Institute for Information Technology, Kyushu University.

## References

[1] Mark D. Allendorf and Robert J. Kee, J. Electrochem. Soc. 138 (1991) 841-852. https://iopscience.iop.org/article/10.1149/1.2085688

[2] Stefano Leone, Olof Kordina, Anne Henry, Shin-ichi Nishizawa, Örjan Danielsson, and Erik Janzén, Crystal Growth & Design 12 (2012) 1977-1984. https://doi.org/10.1021/cg201684e

[3] Tomoya Nagahashi, Hajime Karasawa, Ryota Horiike, Tomoya Kimura, and Kenji Shiraishi, Japanese Journal of Applied Physics, **62** (2023) 048002. https://iopscience.iop.org/article/10.35848/1347-4065/acc3e8

[4] F. Loumagne, F. Langlais, R. Naslain, Journal of Crystal Growth **155** (1995) 205-213. https://www.sciencedirect.com/science/article/pii/0022024895001816

[5] Yuichi Funato, Noboru Sato, Yasuyuki Fukushima, Hidetoshi Sugiura, Takeshi Momose, and Yukihiro Shimogaki, ECS Journal of Solid State Science and Technology **6** (2017) 399-404. https://iopscience.iop.org/article/10.1149/2.0141707jss

[6] Pitsiri Sukkaew, Örjan Danielsson, and Lars Ojamäe, Journal of Physical Chemistry A 122 (2018) 2503-2512. https://doi.org/10.1021/acs.jpca.7b10800

[7] Kenta Chokawa, Yoshiaki Daigo, Ichiro Mizushima, Takashi Yoda, and Kenji Shiraishi, Japanese Journal of Applied Physics, **60** (2021) 085503. https://iopscience.iop.org/article/10.35848/1347-4065/ac1127

[8] Tomoya Kimura, Kenta Chokawa, Kenji Shiraishi, and Atsushi Oshiyama, Physical Review B **106** (2022) 035309.

https://journals.aps.org/prb/abstract/10.1103/PhysRevB.106.035309

[9] S. Yamauchi, I. Mizushima, T. Yoda, A. Oshiyama, K. Shiraishi, Appl. Phys. Express, **17** (2024) 085501. https://iopscience.iop.org/article/10.35848/1882-0786/ad524c

[10] Misagh Ghezellou, Erlend Lemva Ousdal, Marianne E Bathen, Lasse Vines and Jawad Ul-Hassan, Journal of Physics: Materials, **8** (2025) 025008. https://doi.org/10.1088/2515-7639/adb7c0

[11] H. Saitoh, A. Manabe, and T. Kimoto, Materials Science Forum **527–529** (2006) 223–226. https://doi.org/10.4028/www.scientific.net/MSF.527-529.223

[12] U. Forsberg, O. Danielsson, A. Henry, M. K. Linnarsson and E. Janzen, Journal of Crystal Growth **236** (2002) 101–112. https://doi.org/10.1016/S0022-0248(01)02198-4

[13] Gabriel Ferro, Didier Chaussende, Scientific Reports **7** (2017) 43069. https://www.nature.com/articles/srep43069

[14] Kazutoshi Kojima, Satoshi Kuroda, Hajime Okumura, and Kazuo Arai, Applied Physics Letters, **88** (2006) 021907. https://doi.org/10.1063/1.2164912

[15] Tsunenobu Kimoto, Akira Itoh, and Hiroyuki Matsunami, Applied Physics Letters **67** (1995) 2385. https://doi.org/10.1063/1.114555

[16] Takashi Matsuoka and Yoshihiro Kangawa (Eds.), Epitaxial Growth of III-Nitride Compounds: Computational Approach, Springer Series in Materials Science, vol. 269, Springer Cham, 2018. https://doi.org/10.1007/978-3-319-76641-6

[17] Y. Kangawa, A. Kusaba, T. Akiyama, S. Nitta, M. Iwaya, H. Miyake, P. Kempisty, M. A. Chabowska, H. Popova, M. A. Załuska-Kotur, Appl. Phys. Express (NEXT), **19** (2026) 010101. https://iopscience.iop.org/article/10.35848/1882-0786/ae2f19

[18] Y. Kangawa, T. Akiyama, T. Ito, K. Shiraishi, T. Nakayama, Materials **6** (2013) 3309-3360. https://www.mdpi.com/1996-1944/6/8/3309

[19] Y. Kangawa, T. Ito, A. Taguchi, K. Shiraishi, T. Ohachi, Surface Science, **493** (2001) 178-181. https://www.sciencedirect.com/science/article/pii/S0039602801012109

[20] Y. Kangawa, T. Ito, Y. S. Hiraoka, A. Taguchi, K. Shiraishi, T. Ohachi, Surface Science, **507** (2002) 285-289. https://www.sciencedirect.com/science/article/pii/S0039602802012591

[21] K. Momma, F. Izumi, VESTA 3 for three-dimensional visualization of crystal, volumetric and morphology data. J. Appl. Crystallogr. **44** (2011) 1272－1276. https://doi.org/10.1107/S0021889811038970

[22] G. Kresse, J. Furthmü, Efficient iterative schemes for ab initio total-energy calculations using a plane-wave basis set, Phys. Rev. B, **54** (1996) 11169. https://doi.org/10.1103/PhysRevB.54.11169

[23] J. P. Perdew, K. Burke, M. Ernzerhof, Generalized Gradient Approximation Made Simple, Phys. Rev. Lett. **77** (1996) 3865. https://doi.org/10.1103/PhysRevLett.77.3865

[24] P. E. Blochl, Projector augmented-wave method, Phys. Rev. B, **50** (1994) 17953. https://doi.org/10.1103/PhysRevB.50.17953

[25] G. Kresse, D. Joubert, 'From ultrasoft pseudopotentials to the projector augmented-wave method', Phys. Rev. B, **59** (1999) 1758. https://doi.org/10.1103/PhysRevB.59.1758

[26] Hendrik J. Monkhorst and James D. Pack, Special points for Brillonin-zone integrations, Physical Review B, **13** (1976) 5188. https://doi.org/10.1103/PhysRevB.13.5188

[27] Ask Hjorth Larsen, Jens Jørgen Mortensen, Jakob Blomqvist, Ivano E Castelli, Rune Christensen, Marcin Dułak, Jesper Friis, Michael N Groves, Bjørk Hammer, Cory Hargus, Eric D Hermes, Paul C Jennings, Peter Bjerre Jensen, James Kermode, John R Kitchin, Esben Leonhard Kolsbjerg, Joseph Kubal, Kristen Kaasbjerg, Steen Lysgaard, Jón Bergmann Maronsson, Tristan Maxson, Thomas Olsen, Lars Pastewka, Andrew Peterson, Carsten Rostgaard, Jakob Schiøtz, Ole Schütt, Mikkel Strange, Kristian S Thygesen, Tejs Vegge, Lasse Vilhelmsen, Michael Walter, Zhenhua Zeng and Karsten W Jacobsen, Journal of Physics: Condensed Matter, **29** (2017) 273002. https://iopscience.iop.org/article/10.1088/1361

-648X/aa680e

[28] Akira Kusaba, Yoshihiro Kangawa, Pawel Kempisty, Hubert Valencia, Kenji Shiraishi, Yoshinao Kumagai, Koichi Kakimoto and Akinori Koukitu, Japanese Journal of Applied Physics, **56** (2017) 070304. https://iopscience.iop.org/article/10.7567/JJAP.56.070304